\RequirePackage{fix-cm}

\documentclass[smallextended]{svjour3}       
\smartqed  
\usepackage{graphicx}
\begin{document}

\title{Theory of adiabatic fountain resonance
}


\author{Gary A. Williams}

\institute{Gary  A. Williams \at
             University of California, Los Angeles, CA 90095 USA \\  
              \email{gaw@ucla.edu}            
}

\date{}

\maketitle

\begin{abstract}
The theory of ``Adiabatic Fountain Resonance" with superfluid $^4$He is clarified. In this geometry a film region between two silicon wafers bonded at their outer edge opens up to a central region with a free surface. We find that the resonance in this system is not a Helmholtz resonance as claimed by Gasparini and co-workers, but in fact is a fourth sound resonance. We postulate that it occurs at relatively low frequency because the thin silicon wafers flex appreciably from the pressure oscillations of the sound wave.
\keywords{superfluid acoustics \and adiabatic fountain resonance \and fourth sound}
 \PACS{67.25.dg \and 67.25.bh \and 67.25.dt}
\end{abstract}

A low-frequency resonance was observed a number of years ago in superfluid helium by Gasparini and coworkers \cite{g1,g2}.  Their geometry, shown in Fig. 1, consists of a thin helium film sandwiched between two cylindrical silicon wafers bonded together at their outer edges, and opening up to a central channel filled with bulk helium having a free surface.  The resonance was generated with an oscillating heat source, and detected with a thermometer.  A theory of the resonance was subsequently developed, claiming that it is a Helmholtz resonance, with the restoring force being the compressibility of the helium.  Since the bulk liquid surface will oscillate as liquid flows in and out of the film region, the authors named this an ``Adiabatic Fountain Resonance" (AFR).  A number of further papers using this mode to investigate various properties of the helium film have been published \cite{g3,g4,g5,g6,g7,g8,g9,g10,g11,g12,g13,g14},

We argue in this paper that the resonance mode has been misidentified, that in fact no such Helmholtz mode can exist in the configuration of Fig.\,1.  It is proposed instead that the mode is a fourth sound resonance, with an unusually low frequency because the mode is at least partially pressure-released due to the flexibility of the very thin silicon wafers.

In analyzing the theory of the mode, Gasparini and coworkers correctly formulated the two-fluid differential equations involving the superfluid acceleration and mass conservation, Eqs.\,4 and 14 of Ref.\,\cite{g2}.  However, instead of solving these equations, they then made an approximation of setting spatial derivatives equal to constants.  With only time derivatives remaining, the equations reduce to a simple harmonic oscillator of angular frequency
\begin{equation}
\omega _0^2 = \frac{{{\rho _{sc}}}}{\rho }\frac{\sigma }{{\rho \,V\;{l_p}({K_T} + K_C)}}
\end{equation}
where ${\rho _{sc}}$ and $V$ are the superfluid density and volume of the confined helium, $\sigma$ the area of the opening to the bulk, $K_T$ the helium compressibility and $K_C$ the compressibility of the silicon wafers, and ${l_p}$ the arbitrary length resulting from approximating the spatial derivatives.   In this mode  the  the radial superfluid velocity  is found to vary as $1/r$, the result of mass conservation  in the cylindrical geometry (Eq.\,14 of Ref.\,\cite{g2}). The $1/r$ behavior means that at the outer closed wall ($R$ = 2.2 cm in the experiments) there will be finite superflow into the wall.  This cannot exist: any correct solution of the hydrodynamics must have zero flow into a wall.

\begin{figure}
\begin{center}
\includegraphics[width=1.0\textwidth]{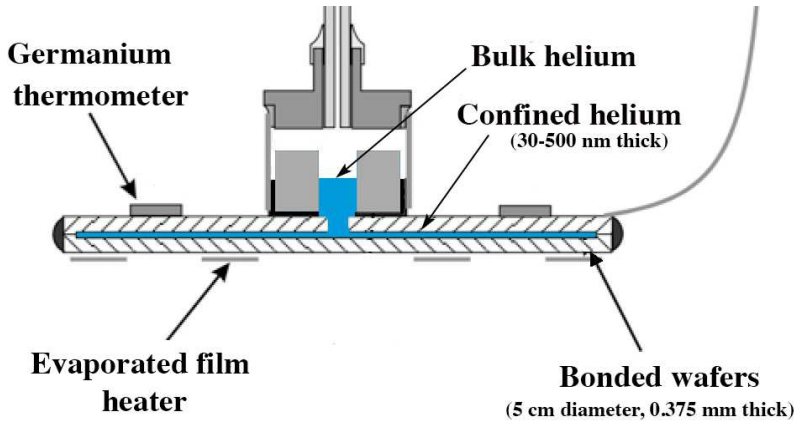} 
\end{center}
\caption{Cell used for Adiabatic Fountain Resonance measurements, after Ref.\, \cite{g2}.  (Color figure online.)}
\label{fig:1}       
\end{figure}

The differential equations cited above in fact are well known \cite{atkins,rudnick}  to have an exact solution if the spatial derivatives are not set constant.  Combined with the compressibility equation of state of the system, the solution is easily found to result in a radial wave equation, with a sound velocity given by adiabatic fourth sound,
\begin{equation}
c_4^2 = \frac{{{\rho _{sc}}}}{\rho }\left( {\frac{1}{{\rho \;({K_T} + {K_C})}}} \right) + c_5^2
\end{equation}
where the fifth sound velocity \cite{gwfifth,maynard,fifth} is
\begin{equation}
c_5^2 = \frac{{{\rho _{nc}}}}{\rho }c_2^2 
\end{equation}
where $c_2$ is the second sound velocity.  The fifth sound term arises from the restoring forces induced by the temperature gradients in the equation for the superfluid acceleration, while the first sound term comes from the pressure gradients.
 
For the cylindrical geometry of Fig.\,1 the solution of the wave equation for the oscillating pressure is a linear combination of the Bessel functions $J_0$ and $Y_0$.  By applying boundary conditions that $v_s = 0$ at the outer wall of radius $R$, and approximating the pressure to be zero at the opening to the bulk fluid at the inner radius $r_0$, the possible wavenumbers $k = 2 \pi / \lambda$ are found by solving 
\begin{equation}
{J_1}\left( {kR} \right) =  - \frac{{{J_0}\left( {k{r_0}} \right)}}{{{Y_0}\left( {k{r_0}} \right)}}{Y_1}\left( {kR} \right)
\end{equation}
Since $r_0$ is typically quite small (0.025 cm), a good approximation to this is ${J_1}\left({kR} \right) = 0$, giving for the lowest mode a wavenumber $k = 3.83 / R$, or a frequency $f = 3.83 \,c_4 / 2\pi R$.  Comparing this to the frequencies reported in Ref.\,\cite{g2} for the cell with thickness 48.3 nm at temperatures 1.85 and 2.152 K gives an effective first sound velocity 
\begin{equation}
({c_1})_{eff} = \sqrt {\frac{1}{{\rho ({K_T} + {K_C})}}}  \approx 35\;m/s  .
\end{equation}
From the known first sound speed the compressibility of the cell is then determined as $K_C \approx 42\, K_T$.  The cell compressibility $K_C = \Delta V / (V \Delta p)$ is appreciable even though there are bonded support posts every 200 $\mu$m throughout the cell.  This results because the cell volume is very small ($V \approx 7\times10^{-11}$ m$^3$ for the 48.3 nm cell), and the wafers are very thin (375 $\mu$m), giving $\Delta V / \Delta p \approx 4 \times 10^{-16}$ m$^3$/Pa, which 
will vary as the inverse cube of the wafer thickness.  For a typical sound amplitude of $\Delta p \approx$ 0.1 Pa this would mean an average wafer deflection of only 0.03 picometers gives the above degree of pressure release.

\begin{figure}
\begin{center}
\includegraphics[width=1.0\textwidth]{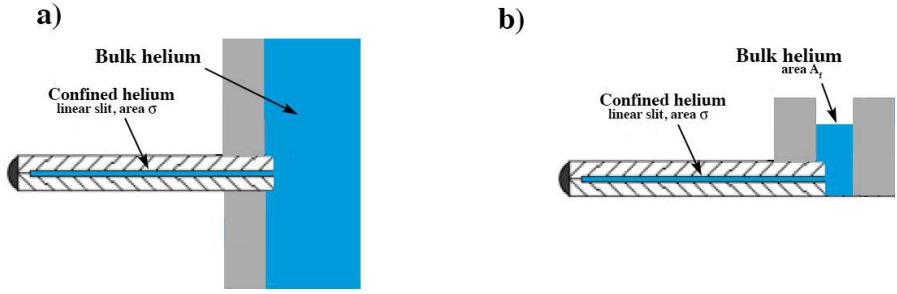} 
\end{center}
\caption{a) Linear slit resonator of length $L$ open at one end to bulk helium.  b) Linear slit resonator opening to bulk helium with a free surface (Color figure online.)}
\label{fig:2}       
\end{figure}
It is simpler to consider the linear equivalent of the cylindrical AFR resonator, shown in Fig.\,2a.  This would be a rectangular slit resonator of uniform cross-section 
$\sigma$ and length $L$, opening to a helium bath at one end and closed at the other end.  The differential equations for the superfluid acceleration and mass conservation are still given by Eqs.\,4 and 14 of Ref.\,\cite{g2}, and if we make the same unjustified approximation of setting the spatial gradient terms equal to a constant, the resonant frequency is found to be given again precisely by Eq.\,1.  However, it is now even easier to see that this is a completely spurious mode:  
in Eq.\,14 of Ref.\,\cite{g2} the cross-section $\sigma$ will now be a constant rather than varying as $r$ in the cylindrical geometry.  Conservation of mass thus requires that the superfluid velocity be a constant over the entire channel, from the inlet flow to the solid end wall.  Such hydrodynamic flow into a wall is simply impossible.  

If the spatial gradients are not set constant for the linear resonator, there is again an exact solution, a wave equation with wave speed given by Eq.\,2.   Boundary conditions on the wave solutions are zero superfluid velocity at the closed end, while at the open end a more rigorous condition than zero pressure is needed to account for the sound generated into the bulk region, arising from the motion of the superfluid in and out of the channel end.  It is well known that this ``end effect" \cite{kinsler} leads to an increase in the effective length of the channel, to $L + \Delta L$.  
The lowest mode frequency of this closed-open tube is then the usual quarter-wave resonance
\begin{equation}
f _1= \frac{{{c_4}}}{{4(L + \Delta L)}}   .
\end{equation}
The calculation of $\Delta L$ differs from the acoustics textbook calculation \cite{kinsler} in two aspects:  the superfluid density $\rho_{sc}$ in the confined helium in the channel is not necessarily the same as the bulk superfluid density $\rho_{sb}$, and both first and second sound will be generated in the bulk helium, giving 
$\Delta L = \Delta L_1 + \Delta L_2$.
From the first sound generation an approximate calculation gives
\begin{equation}
\Delta L_1 = \frac{{{\rho _{sb}}}}{{{\rho _{sc}}}}\frac{8}{{3\pi }}\sqrt {\frac{\sigma }{\pi }}
\end{equation}
where $\sigma$ is the cross-sectional area of the slit opening.  $\Delta L_2$ has not been calculated, but we expect it to be similar in magnitude to $\Delta L_1$.  $\Delta L$ is typically a small correction to $L$, but note that if $\rho_{sc}$ becomes much smaller than $\rho_{sb}$ (as will occur at and above the finite-frequency Kosterlitz-Thouless transition temperature \cite{ahns}) it can become quite large.  In that case nearly all of the kinetic energy will be the flow near the opening, which will dominate the resonant frequency.  

The case shown in Fig.\,2b can also be considered, with the bulk having a free surface.  The flow in and out of the channel will cause the free surface to oscillate in height: a fountain.  This will add an additional gravitational restoring pressure $\delta p = \rho g \delta h$.  In the low-temperature limit where the additional thermal fountain pressure \cite{robinson} can be neglected the lowest mode frequency increases slightly,
\begin{equation}
{f_1} = \frac{{{c_4}}}{{4\;(L + \Delta L)}} + \frac{g}{{\frac{{{\pi ^2}}}{4}\frac{{{\rho _{sb}}}}{{{\rho _{sc}}}}\frac{{{  A_f}}}{\sigma }{c_4}}}
\end{equation}
where $A_f$ is the area of the bulk surface.   This additional term is quite small, except again in a region where $\rho_{sc} << \rho_{sb}$.  There is no additional low-frequency ``gravitational" mode as proposed by Gasparini and coworkers \cite{g2}, since this geometry is quite different from the U-tube mode considered by Robinson \cite{robinson}.

It is certainly possible that the fourth sound mode in these resonators is not completely adiabatic, since the thermal conductivity of silicon is fairly large.  However, the Kapitza thermal resistance \cite{pollack} at the boundary between the helium and the silicon makes it impossible to eliminate thermal gradients in the  helium even if the wall conductivity is high.  The fact that the Gasparini group still observes the temperature oscillations of the fourth sound mode shows that this is the case.  Even if the fourth sound is nearly isothermal (which we doubt) there is no change to our arguments:  there will still be a pressure gradient and zero flow at the wall, and the fourth sound velocity will be only slightly changed, since the fifth sound velocity is still small compared to the effective first sound velocity. 

There is one situation where it is valid to neglect spatial derivatives.  Consider a geometry where the closed end of the linear resonator of Fig.2 is instead opened and positioned upright at the liquid helium surface.  A temperature gradient between bulk and film will then give rise to a true superfluid fountain, with mass conservation guaranteeing a constant superfluid velocity along the channel.  However, if the end is now closed back to a solid wall everything changes.  The new boundary condition of zero mass flow will give rise to pressure and temperature gradients, such that with an oscillatory drive the only possible resonance is a fourth-sound mode.  The same exact arguments in fact apply to a similar classical system: a uniform pipe filled with water and open at both ends.  A pressure gradient will of course produce a constant-velocity fountain, but if the end is now closed off with a solid wall the only possible resonance mode is the quarter-wave resonance of an open-closed pipe, the classical-fluid limit of Eq. 6.  

The explanation in terms of fourth sound solves a strange feature seen in recent measurements of AFR in cells having thousands of additional $\mu$m$^3$ ``boxes" in addition to the flat film \cite{g10,g11,g12,g13}.  In these cells the AFR mode was found to be split into two slightly different frequencies.  This was ascribed to the fact that the boxes did not reach all the way out to the edge of the wafers, with the last few millimeters being just the uniform film.  This is actually impossible to understand with the ``Helmholtz" interpretation of AFR, since Helmholtz modes are completely independent of the details of the cell geometry, depending only on the total volume as in 
Eq.\,1.  In contrast, the fourth sound will be quite sensitive to the geometry, with the mode splitting into two different wavelengths, one resulting from reflection at the outer wall, and the other slightly shorter due to reflections at the boundary between the box region and the uniform film.

In summary, we propose that the AFR mode has a simple explanation as partially pressure-released fourth sound.  This resolves a number of contradictions in the ``Helmholtz" theory.  A more complete analysis of data near the Kosterlitz-Thouless transition will require a formulation of the correction factors in Eqs.\,7 and 8 for the cylindrical geometry.  An interesting application of these nano-patterned cells would be to construct one without bonding the support posts, so that the walls could flex freely, providing nearly complete pressure release to the sound mode, $K_C >> K_T$.  In this limit the mode would be purely fifth sound, allowing a study of this mode without the necessity of first subtracting additional restoring forces such as surface tension or Van der Waals forces \cite{gwfifth,maynard}.

\begin{acknowledgements}
We thank Seth Putterman for useful discussions.
\end{acknowledgements}


\end{document}